\documentclass{ws-mpla}

\usepackage{url}
\usepackage[super]{cite}
\usepackage{amsmath,amsfonts,amssymb,dsfont,mathrsfs,centernot}
\usepackage{graphicx}
\usepackage[colorlinks=True,linkcolor=blue,citecolor=blue]{hyperref}
\usepackage{xcolor}
\usepackage{comment}
\usepackage{siunitx}

\newcommand{\abs}[1]{\left|#1\right|}

\newcommand{\cdf}{{\boldsymbol{\mathcal{D}}}}
\newcommand{\covd}{\mathcal{D}}

\newcommand{\df}{{\bf{d}}}

\newcommand{\fy}{\centernot}

\newcommand{\ga}{\gamma}

\newcommand{\J}{\mathscr{J}}
\newcommand{\Lag}{\mathscr{L}}

\newcommand{\Or}{\mathscr{O}}

\newcommand{\Ri}{\mathcal{R}}

\newcommand{\tor}{\mathcal{T}}
\newcommand{\w}{\wedge}

\newcommand{\bps}{\ensuremath{\bar{\psi}}}
\newcommand{\Bps}{\ensuremath{\bar{\Psi}}}

\newcommand\VIF[1]{\hat{\bf{e}}^{\hat{#1}}}

\newcommand\hvif[1]{\hat{\bf{e}}^{{#1}}}

\newcommand\PA[1]{\partial_{\hat{#1}}}

\newcommand\SPI[1]{\hat{\omega}_{\hat{#1}}}
\newcommand\SPIF[2]{\hat{\boldsymbol{\omega}}^{\hat{#1}}{}_{\hat{#2}}}
\newcommand{\RIF}[2]{\hat{\boldsymbol{\mathcal{R}}}^{\hat{#1}}{}_{\hat{#2}}}
\newcommand{\hRif}[2]{\hat{\boldsymbol{\mathcal{R}}}^{{#1}}{}_{{#2}}}

\newcommand{\TF}[1]{\hat{\boldsymbol{\mathcal{T}}}^{\hat{#1}}}

\newcommand{\hcont}[3]{\hat{\mathcal{K}}_{#1}{}^{#2}{}_{#3}}

\newcommand{\CONTF}[2]{\hat{\boldsymbol{\mathcal{K}}}^{\hat{#1}}{}_{\hat{#2}}}

\newcommand{\beq}{\begin{equation}}
\newcommand{\eeq}{\end{equation}}
\newcommand{\ber}{\begin{eqnarray}}
\newcommand{\eer}{\end{eqnarray}}

\renewcommand{\(}{\left(}
\renewcommand{\)}{\right)}
\renewcommand{\[}{\left[}
\renewcommand{\]}{\right]}

\newcommand{\dn}[2]{\,{\rm{d}}^{#1}{#2}\;}

\newcommand*{\de}[1][]{\mathop{\mathrm{d}#1}\nolimits}% differential, facultative argoment between square parentheses
\hypersetup{%
  pdftitle={Updated Limits   on Extra Dimensions through Torsion and LHC Data},
  pdfauthor={Oscar Castillo-Felisola},
  pdfsubject={High Energy Physics},
  pdfkeywords={Extra dimensions; Torsion; Modified Gravity; LHC},
  pdflang={English}
}

\begin{document}

\markboth{O. Castillo-Felisola, C. Corral, I. Schmidt and A. R. Zerwekh}
{Updated Limits   on Extra Dimensions through Torsion and LHC Data}

%%%%%%%%%%%%%%%%%%%%% Publisher's Area please ignore %%%%%%%%%%%%%%
\catchline{}{}{}{}{}
%%%%%%%%%%%%%%%%%%%%%%%%%%%%%%%%%%%%%%%%%%%%%%%%%%%%%%%%%%%%%%%%%%%

\title{UPDATED LIMITS   ON EXTRA DIMENSIONS THROUGH TORSION AND LHC DATA}

\author{OSCAR {CASTILLO-FELISOLA},\footnote{Corresponding author}\quad CRISTOBAL {CORRAL},}
\author{IV\'AN {SCHMIDT} and ALFONSO R. {ZERWEKH}}
%% \email{alfonso.zerwekh@usm.cl}
\address{Centro Cient\'\i fico Tecnol\'ogico de Valpara\'\i so,\\
  Valparaiso, Chile.\\
  and\\
  Departamento de F\'\i sica, Universidad T\'ecnica Federico Santa Mar\'\i a,\\
  Casilla 110-V,  Valpara\'\i so, Chile.\\
  o.castillo.felisola@gmail.com
  %% \\
  %% cristobal.corral@postgrado.usm.cl\\
  %% ivan.schmidt@usm.cl\\
  %% alfonso.zerweh@usm.cl
}

\maketitle

\pub{Received (Day Month Year)}{Revised (Day Month Year)}

\begin{abstract}
  It is well known that inclusion of torsion in the gravitational formalism, leads to  four-fermion interactions. Although the coupling constant of this interaction is strongly suppressed in four dimensions, its value is enhanced in models with $n$ extra dimensions.  
  In this context, we reinterpret the recent limits established by LHC experiments to four-fermion contact interactions,  to set bounds on the size of the extra dimensions. 
  For  $n=2$, the limits are comparable to those in the literature, while for $n\geq3$ the volume of the extra dimensions is strongly constrained.

\end{abstract}

\ccode{PACS Nos.: 04.50.Kd, 04.62.+v, 11.25.Mj}

\maketitle

\section{Introduction}\label{sec:intro}

The standard model (SM) of particle physics has proved to be a very successful theory, but still has problems. Among these, one that is considered fundamental is the vast difference between the electroweak and gravitational scales, known as the {\it hierarchy problem}, which might be solved introducing extra dimensions. The gravitational theory of Einstein can be generalized in several ways, specially in higher dimensional spacetimes. One of the simplest generalizations is due to Cartan, who proposed that the torsion does not necessarily vanishes.\cite{Cartan1922,Cartan1924,Cartan1925} 

Cartan's extension of general relativity, can be naturally written in the first order formalism, using two independent kind of fields: (a)  the vielbeins, which encode the information carried by the metric, and (b) the spin connection, which encipher the way the parallel transport is performed. However, in general the spin connection possesses an antisymmetric part, that is not present in the Levi-Civita one, which introduces torsion to the picture. In this formalism, gravity can be coupled to fermionic matter  naturally. Since the field equation for the spin connection is algebraic, it can be used to substitute the torsion in terms of the spin of fermion fields. The new action contains standard general relativity and matter fields with an additional contact four-fermion interaction.\cite{Kibble:1961ba,RevModPhys.48.393,Shapiro:2001rz,SUGRA-book,Castillo-Felisola:2013jva}  

In four dimensions, within the most straightforward torsional extension of Einstein-Hilbert gravity,\footnote{Using a non-vanishing torsion, more general  gravitational models have been constructed with terms quadratic in torsion.\cite{PhysRevD.18.2730,PhysRevD.22.1915,Baekler2011}} the effective four-fermion interaction term has a coupling constant proportional to Newton's gravitational constant, $G_N\sim M_{\text{pl}}^{-2}$, where $M_{\text{pl}}$ is the Planck's mass; although in the most general torsional generalization of Einstein gravity, the effective four-fermion interaction term has a coupling constant proportional to a yet undetermined constant.\footnote{As remarked by L. Fabbri, the coupling constant of the effective four-fermion interaction term, generically differs from the Newton's gravitational constant.\cite{Fabbri:2011kq}} Therefore, at first glimpse this interaction is highly suppressed. Nevertheless, in the last twenty years diverse scenarios have proposed that the existence of extra dimensions could solve the hierarchy problem, due to the introduction of a (higher dimensional) fundamental gravity scale of  roughly $M_*\sim \Or(1) \;\si{TeV}$.\cite{ADD1,ADD2,AADD,RS1,RS2} Recently, limits to the fundamental scale of gravity  have been set up  by direct searches of 
quantum black holes\cite{Gingrich:2012vb} and the influence of the exchange of virtual gravitons on dilepton events\cite{Aad201240,Aad:2012bsa}.
These might be compared with  gravitational theories with torsion through their observables torsional observables\cite{RevModPhys.48.393,Shapiro:2001rz,Belyaev:1998ax,PhysRevD.75.034014}, including some generalisations to higher dimensional spacetimes\cite{Chang:2000yw,Lebedev:2002dp} or Lorentz violating models.\cite{Kostelecky:2007kx}  %have been study previously, 

%%%%%%%%% Add references about direct or indirect searches of torsion and current limits %%%%%%%%%
%% \cite{Khriplovich:2012yj} Cosmology constrains gravitational four-fermion interaction
%% \cite{Chang:2000yw} Universal Torsion–Induced Interaction from Large Extra Dimensions
%% \cite{Lebedev:2002dp} Torsion constraints in the Randall-Sundrum scenario
%% \cite{Belyaev:1998ax,PhysRevD.75.034014} Torsion action and its possible observables; Torsion phenomenology at the CERN LHC
%% \cite{Kostelecky:2007kx} New Constraints on Torsion from Lorentz Violation

On the other hand,  ATLAS and CMS collaborations have presented experimental limits for the coupling constant of four-fermion contact interaction.\cite{Aad:2011aj,Aad201240,ATLAS:2012pu,Aad:2012bsa,PhysRevD.87.032001,Chatrchyan:2013muj} 
Adequately interpreted, these results are useful for imposing bounds on the value of the fundamental gravity scale, $M_*$, in the context of theories with torsion. By extension, it is possible to find limits on the size of the eventual extra dimensions.
%%%%%%%%% References of the limits, what are we willing to do and how our work differs from the previous results %%%%%%%%%

The aim of this work is to obtain updated  bounds and limits on the typical size of the extra dimensions using data coming from the LHC experiment.\footnote{Constraint to four-fermion interactions have been found from cosmological data,\cite{Khriplovich:2012yj} but here cosmological data have not been considered.}  The paper is organised as follows, in Sec.~\ref{sec:CEF}, for the sake of completeness,  a brief presentation of Cartan's generalisation of gravity coupled with fermions is presented. In the next sections constraints are found for different higher dimensional scenarios, such as those proposed by Arkani-Hamed, Dimopoulos and Dvali\cite{ADD1,ADD2,AADD} in Sec.~\ref{sec:ADD}, and by Randall and Sundrum\cite{RS1,RS2} in Sec. \ref{sec:RS}.  In Sec.~\ref{sec:res} a summary of results and conclusions is presented.

\section{Cartan-Einstein Gravity with Fermions}\label{sec:CEF}

It is a well-known fact that Einstein's gravity is a field theory for the metric, and the spacetime connection is required to be torsion-free (see for example Ref.~\refcite{Wald:book}). 

Interestingly, when the first order formalism of pure gravity is considered, \textit{i.e.} with all matter fields turned off, whose fields are the vielbeins and spin connection, the torsion-free imposition is nothing but the equation of motion for the spin connection.\cite{Cartan-Einstein}

However, for gravity coupled with fermionic matter, this condition changes and introduces a four-fermion contact interaction. The modification of the fermionic Lagrangian due to the presence of torsion is presented below, using the notation in Ref.~\refcite{Castillo-Felisola:2013jva}.

The Cartan-Einstein action in $D$-dimensions is,
\begin{align}
  S_{\text{gr}} = \frac{1}{2\kappa^2}\int\frac{\epsilon_{\hat{a}_1\cdots \hat{a}_D}}{(D-2)!}\hRif{\hat{a}_1 \hat{a}_2}{}\w\hvif{\hat{a}_3}\w\cdots\w\hvif{\hat{a}_D},\label{CE-action}
\end{align}
where $\VIF{a}$ and $\RIF{a}{b}$ are the vielbeins 1-form and curvature 2-forms respectively, and these are related with the spin connection 1-form, $\SPIF{a}{b}$, and torsion 2-form, $\TF{a}$,\footnote{The curvature and torsion 2-forms are related with the usual curvature and torsion tensors through the relations, $\hRif{\hat{a}\hat{b}}{}=\frac{1}{2}\hat{\Ri}^{\hat{a}\hat{b}}{}_{\hat{\mu}\hat{\nu}}\,\de[x]^{\hat{\mu}}\w\de[x]^{\hat{\nu}}$, and $\TF{a}=\frac{1}{2} \hat{\mathcal{T}}_{\hat{\mu}}{}^{\hat{a}}{}_{\hat{\nu}}\,\de[x]^{\hat{\mu}}\w\de[x]^{\hat{\nu}}$.} through the so called structural equations,
\begin{align}
  \df\VIF{a}+\SPIF{a}{b}\w\VIF{b} &= \TF{a},\label{struc.eq.1}\\
  \df\SPIF{a}{b}+\SPIF{a}{c}\w\SPIF{c}{b}&=\RIF{a}{b}.\label{struc.eq.2}
\end{align}
Additionally, the Dirac action in arbitrary dimension is
\begin{align}
  S_\Psi ={}& -\int \frac{\epsilon_{\hat{a}_1\cdots \hat{a}_D}}{(D-1)!} \Bps \hvif{\hat{a}_1}\w\cdots\w\hvif{\hat{a}_{D-1}}\ga^{\hat{a}_D}\hat{\cdf}\Psi \notag\\
 & -m\int\frac{\epsilon_{\hat{a}_1\cdots \hat{a}_D}}{D!}\Bps \hvif{\hat{a}_1}\w\cdots\w\hvif{\hat{a}_{D}}\Psi,
\end{align}
with $\hat{\cdf}$ the exterior derivative twisted by the spin connection, $$\hat{\cdf}= \de[x]^{\hat{\mu}}\hat{\covd}_{\hat{\mu}}= \de[x]^{\hat{\mu}}\PA{\mu} + \frac{\imath}{2} \de[x]^{\hat{\mu}}\(\SPI{\mu}\)^{\hat{a}\hat{b}}\J_{\hat{a}\hat{b}}, $$ with $\J_{\hat{a}\hat{b}}$ the generators of the spinorial representation of the Lorentz group.

Since the total action is 
\begin{align}
  S = S_{\text{gr}} + S_{\Psi},
\end{align}
the equations of motion for the whole system are,\footnote{Multi-indices $\ga$'s denote the anti-symmetric product of $\ga$-matrices.}
\begin{align}
  \hat{\Ri}^{\hat{m}}{}_{\hat{a}_3} -\frac{1}{2}\hat{\Ri} \delta^{\hat{m}}_{\hat{a}_3} &=\kappa^2\bar{\Psi}\[\ga^{\hat{m}}\hat{\covd}_{\hat{a}_3}-\delta^{\hat{m}}_{\hat{a}_3}\(\fy{\hat{\covd}}+m\)\]\Psi, \label{ED-eom}\\
  %\hcont{\hat{a} \hat{b} \hat{c} }{}{} \equiv
  \frac{1}{2}\(\hat{\tor}_{\hat{b} \hat{c} \hat{a} } + \hat{\tor}_{\hat{b} \hat{a} \hat{c} } + \hat{\tor}_{\hat{a} \hat{b} \hat{c} }\) &= -\frac{\kappa^2}{4} \bar{\Psi}\ga_{\hat{a} \hat{b} \hat{c}}\Psi, \label{cont-constraint}
\end{align}
the LHS of Eq.~\eqref{cont-constraint} is generally denoted by  $\hcont{\hat{a} \hat{b} \hat{c} }{}{}$, and is called  contorsion. Notice that in the case of completely antisymmetric spin density, as in the case of Dirac fields above, it is enough to consider the completely antisymmetric part of the torsion (as in Eq.~\eqref{cont-constraint}) in what will be the single squared-torsion contribution.\cite{Fabbri:2012yg} 

Since Eq.~\eqref{cont-constraint} is a constraint, it can be substituted back into the action. %, using that
It is done by splitting  the spin connection into a torsion-free part, $\hat{\overline{\boldsymbol{\omega}}}^{\hat{a}}{}_{\hat{b}}$, and  the contorsion. In differential forms, it is expressed as\footnote{The contorsion one-form is obtaines from the contorsion by contracting its first index with the one-form basis, $\CONTF{a}{b}=\de[x]^\mu \hcont{\mu}{\hat{a}}{\hat{b}}$.}
\begin{align}
  \SPIF{a}{b} \mapsto %\SPIF{a}{b}
  \hat{\overline{\boldsymbol{\omega}}}^{\hat{a}}{}_{\hat{b}}+\CONTF{a}{b}.
\end{align}
%where the former is a general spin connection, while the later is a torsion-free one.

The new action, expressed in terms of torsion-free quantities,  is
\begin{align}
  S ={}& \int\dn{D}{x}\abs{\hat{e}}\[\frac{1}{2\kappa^2}\hat{\Ri} -\Bps\(\fy{\hat{\covd}}+m\)\Psi\right.\notag\\
  & \left.+\frac{\kappa^2}{32}\bar{\Psi}\ga_{\hat{a} \hat{b} \hat{c}}\Psi\bar{\Psi}\ga^{\hat{a} \hat{b} \hat{c}}\Psi\],
\end{align}
which is the usual torsion-free theory of gravity coupled to a fermion field with a four-fermion contact interaction.

In the next sections the gravitational effects, other than the generation of a four-fermion interaction, are neglected.
%aspects of this model are not considered, based in the fact that the four-dimensional universe is essentially flat, and gravitational forces are extremely weak in comparison to other known interactions. Nonetheless, the existence of torsion still leaves behind a four-fermion interaction.
Hereon, our objective will be to compare the four-fermion interaction coming from this gravity model with the four-fermion interaction limits found at the LHC  experiments.

%%%\section{Four-Dimensional Effective Theory\label{sec:eff}}
% Do we need to include a section about the dimensional reduction?

\section{Bounds on Four-Fermion Interaction}\label{sec:4fi}

Early proposals of contact four-fermion interaction signals in colliders are found in \cite{RevModPhys.56.579,RevModPhys.58.1065,Chiappetta:1990jd}. These works inspired searches at the LHC experiments. In particular,  ATLAS and CMS  collaborations have found limits on the scale of four-fermion contact interaction, by analyzing the invariant mass and angular distribution of dijets\cite{Aad:2011aj,Aad201240,ATLAS:2012pu,Aad:2012bsa,PhysRevD.87.032001,Chatrchyan:2013muj}. The strongest constraint comes from an interaction of chiral fermions in the form
\begin{align}
  \Lag_{qqqq} = \pm\frac{g^2}{2\Lambda^2}\; \bps_{qL}\ga^a\psi_{qL}\bps_{qL}\ga_a\psi_{qL},\label{Lag-qqqq}
\end{align}
which is experimentally excluded for $\Lambda<\SI{14}{TeV}$, assuming that the coupling constant $g\sim\Or(1)$. 

In this section the above limit is compared with the  four-fermion interaction coming from the Cartan-Einstein gravity action,\footnote{Note that four-dimensional fermions are denoted with lower case symbol $\psi$, while the fermion in arbitrary dimension is denoted with the capitalized one, $\Psi$.} 
\begin{align*}
  \Lag_{4\Psi} = \frac{\kappa^2}{32}\; \Bps\ga_{\hat{a} \hat{b} \hat{c}}\Psi \Bps\ga^{\hat{a} \hat{b} \hat{c}}\Psi.
\end{align*}
Within the context of the straightforward torsional  generalization of Einstein's gravity in four dimensions,\footnote{In our notation, the four-dimensional chiral matrix is called $\ga^*$.} where $\kappa^2\sim M_{\text{pl}}^{-2}$, $\ga_{abc}\sim \ga_d\ga^*$ and $\Psi = \psi_L +\psi_R$,  the comparison implies $\Lambda\sim M_{\text{pl}}\ggg \SI{10}{TeV}\;$, which is tautological.

Nonetheless, there exist models where the fundamental scale of gravity is not $M_{\text{pl}}\sim 10^{18}\;\si{GeV}$, but rather a much lower one, $M_*$, which could be of order of the electro-weak scale, i.e. $M_{*}\sim M_{EW}$, giving a natural solution to the {\it Hierarchy Problem}. These models, however,  require  extra dimensions. Therefore, in the following the spacetime will be considered to be $(4+n)$-dimensional, where the $n$ extra dimensions are either compact or extended, and differentiation between models is made according to this characteristic. %In order to maintain a consistent notation, the distinction  between flat and curved coordinates is though Latin and Greek indices. Moreover, hated indices run over the whole spacetime whilst unhated ones run over the four-dimensional restriction.

In order to achieve the goal of comparison, it is necessary to reduce the dimension of the spacetime from $D$ down to four. Although the dimensional reduction could be a complex procedure,\footnote{The difficulty for finding the effective theory comes from the fact that in higher dimensional spacetimes, the spinorial representation of the Lorentz group could have dimension different than four. Therefore, one should be aware of the decomposition of the spinors (including the profiles through the extra dimensions), as well as the Clifford algebra elements, prior to the integration of the extra dimensions.} we will sketch how a term like Eq.~\eqref{Lag-qqqq} appears. %% In the following, the higher dimensional spacetime would is considered to be five-dimensional.

First, in the contraction $(\ga_{\hat{a} \hat{b} \hat{c}})(\ga^{\hat{a} \hat{b} \hat{c}})$ one splits the indices to consider the four-dimensional contribution,\cite{Castillo-Felisola:2013jva} i.e., $$(\ga_{\hat{a} \hat{b} \hat{c}})(\ga^{\hat{a} \hat{b} \hat{c}}) = (\ga_{abc})(\ga^{abc}) + 3(\ga_{ab*})(\ga^{ab*}) + \cdots ,$$where the ellipsis stand for additional terms, but the focus will be on the first one, because it will rise the kind of interaction comparable with the experimental data. Second, the antisymmetric product of three elements of the four-dimensional Clifford algebra is equal to the product of the missing element of the Clifford algebra times the chiral element, i.e., $\ga_{abc} \sim \ga_d \ga^*$. Next, the higher dimensional spinor $\Psi$ can be decomposed (or compactified) as the product of the four-dimensional times $n$-dimensional spinors,\cite{GSW2,PopeKK,IAS2,CastilloFelisola:2010xh,CastilloFelisola:2012ez}
\begin{align}
  \Psi(x,\xi) &= \sum_i \psi^{(i)}(x)\otimes \lambda_i(\xi) \notag \\
  &= \sum_i \(\psi^{(i)}_L(x) +\psi^{(i)}_R(x)\)\otimes \lambda_i(\xi),\label{fermdecomp}
\end{align}
where the standard Kaluza-Klein decomposition for fermionic fields has been used,\footnote{In general the Kaluza-Klein decomposition in term of the Weyl spinors, accept different \textit{profiles} ($\lambda_L$ and $\lambda_R$). Nonetheless, for the purpose of comparison with the term in Eq.~\eqref{Lag-qqqq} (where all spinors are left-handed), we restrict ourselves to the case of equal profiles.} $x$ denotes the coordinates on the four-dimensional spacetime, $\xi$ denotes the extra dimensional coordinates, and also  Dirac spinors (in four dimensions) have been decomposed in terms of  Weyl spinors.
Therefore, after integration of the extra dimensions, the effective four-dimensional theory would have a term of the desired form
\begin{align}
  \Lag_{\text{eff}} = \frac{\kappa^2_{\text{eff}}}{32}\; \bps_{qL}\ga^a\psi_{qL}\bps_{qL}\ga_a\psi_{qL}.\label{Lag-eff}  %% \; \Bps\ga_{\hat{a} \hat{b} \hat{c}}\Psi \Bps\ga^{\hat{a} \hat{b} \hat{c}}\Psi.
\end{align}  %, say Eq.~\eqref{Lag-qqqq}.
Dimensional  analysis gives two possibilities. Either the effective coupling constant is directly related to the fundamental gravitational scale, $\kappa_{\text{eff}}\sim M_*^{-1}$, or is related to the effective four-dimensional one $\kappa_{\text{eff}}\sim \(M'_{\text{pl}}\)^{-1}$, where $M'_{\text{pl}}$ is the redefined Planck mass after the dimensional reduction. These two interpretations originate different limits, which rise the bound limits presented below. %Since the latter is more restrictive, the size of the extra dimensions would be smaller, $R_{min}$, while the former gives the maximum size of the extra dimensions, $R_{max}$.
In the following, the limits corresponding to the first interpretation will be called $R_{\text{A}}$, while the limit given by the second interpretation will be called $R_{\text{B}}$

\subsection{ADD models}\label{sec:ADD}

ADD models\cite{ADD1,ADD2}  consist of a four-dimensional spacetime with a set of $n$  compact extra dimensions, with typical length $R$. Matter is confined to the four-dimensional spacetime for energies below $\Lambda\sim \frac{1}{R}$, while gravity propagates through the whole spacetime. This configuration allows to solve the hierarchy problem, because the natural scale for gravity is not the effective four-dimensional one, but rather the $(4+n)$-dimensional.

The relation between the fundamental gravitational scale, $M_*$, and the four-dimensional effective one, $M_{\text{pl}}$ is given by
\begin{align}
  M_{\text{pl}}^2 \sim M_*^{2+n} R^n.\label{rel-ADD}
\end{align}
Additionally, the coupling constant of the Einstein action is  $\kappa^2\sim \frac{1}{M_*^{2+n}}$. It is worthwhile to mention that ADD scenarios are restricted to $n\geq 2$, because of gravitational phenomenology.\cite{ADD1}

From Eq.~\eqref{rel-ADD}, it follows that the typical radii of the extra dimensions  are
\begin{align}
  R\sim 10^{\frac{30}{n}-17}\(\frac{\SI{1}{TeV}}{M_*}\)^{\frac{2}{n}+1}\;\si{cm}.
\end{align}
Assuming that the scale of the four-fermion interaction, $\Lambda$, is essentially the fundamental scale of gravity, $M_*$, one finds sizes of the extra dimensions from roughly a few micrometers down  to  a few tens femtometers, depending on the number of extra dimensions considered, as shown in the second column of Table~\ref{Tab:ADD-R}. These results are a refinement of the original ADD claim.
\begin{table}[ht]
  \tbl{Typical radius of the extra dimensions in ADD models. $R_{\text{A}}$ for $\Lambda\sim M_*$. $R_{\text{B}}$ for $\kappa^2_{\text{eff}}\sim \Lambda^{-2}$, assuming $M_*=\SI{100}{GeV}$.}{
    \label{Tab:ADD-R}
    %\begin{center}
      \begin{tabular}{@{}>{$}c<{$\hspace{15mm}} >{$}c<{$\hspace{15mm}} >{$}c<{$} @{}}
        \toprule
        n & R_{\text{A}}[\si{m}]& R_{\text{B}}[\si{m}]\\
        \colrule
        2& 10^{-6} &  10^{-16}\\
        3& 10^{-11}&  10^{-16}\\
        4& 10^{-13}&  10^{-17}\\
        5& 10^{-15}&  10^{-17}\\
        6& 10^{-16}&  10^{-17}\\
        7& 10^{-16}&  10^{-17}\\
        \botrule
      \end{tabular}
  %\end{center}
  }
\end{table}

On the other hand, the effective coupling constant in four dimensions, $\kappa_{\text{eff}}^2$, should be related directly with the fundamental scale of gravity, i.e., $\kappa^2_{\text{eff}}\sim M_*^{-2}$, or in a similar way as before
\begin{align}
  R\sim \(\frac{\Lambda}{M_*}\)^{\frac{2}{n}}\frac{1}{M_*}.
\end{align}
Hence, one finds another limit for the typical size of the extra dimensions, as shown on the third column of Table~\ref{Tab:ADD-R}.

%% One might note that $R_{B}$ does not vary strongly as $n$ changes. Therefore, 
 
\subsection{Randall-Sundrum models}\label{sec:RS}

When considering Randall-Sundrum  brane-worlds scenarios,\cite{RS1,RS2} with metric
\begin{align}
  \hat{g}_{\hat{\mu}\hat{\nu}} = e^{-2k r_c\abs{\xi}}\eta_{\mu\nu}\delta^\mu_{\hat{\mu}}\delta^\nu_{\hat{\nu}} + r_c^2 d\xi_{\hat{\mu}}d\xi_{\hat{\nu}},
\end{align}
the relation between the four-dimensional Planck mass, $M_{\text{pl}}$, and the fundamental (five-dimensional) Planck scale, $M_*$, is given by
\begin{align}
  M_{\text{pl}}^2 = \frac{M_*^3}{k}.%\label{rel-RS}
\end{align}
A well-known modulus stabilization method would ensure that the product $k r_c\sim 10$\cite{Goldberger:1999uk}, the relation between the gravitational scales and the length of the extra dimension is found to be
\begin{align}
  M_{\text{pl}}^2 \sim \frac{M_*^3 r_c}{10}.\label{rel-RS}
\end{align}

Analyzing both limits as in the previous section, the limit on the extra dimension size is
\begin{equation}
  \begin{split}
    R_{\text{A}}&<\SI{e10}{m},\\
    R_{\text{B}}&<\SI{e-13}{m}.
  \end{split}\label{RSlim}
\end{equation}
The range is particularly  wide because there is a single extra dimension. Although brane-worlds of codimension higher than one have been considered\cite{Carroll:2003db,Parameswaran:2006db,Burgess:2008ka,Bayntun:2009im,Nierop-PhD}, without a carefully thought-out moduli stabilization process the bounds on the extra dimensions sizes are equal to those found in Sec.~\ref{sec:ADD} (shown in Table~\ref{Tab:ADD-R}).
%% Add more references with codimension higher than 1

%\section{Results and Conclusions}\label{sec:res}

\section{Concluding Remarks}\label{sec:res}

In this work we have used the limits on four-fermion chiral contact interactions obtained by the LHC collaborations in order to constrain the typical size of eventual extra dimensions in models where gravity admits torsion in the bulk.

Depending of the approach, there are two types of interpretations. These approaches, that we refer to as type A and type B, yield to constraints differ by at least two orders of magnitude. According to the Table~\ref{Tab:ADD-R}, and the limits of Eq.~\eqref{RSlim}, it is more likely to rule out the type B interpretation.

For a codimension 2 we have found an upper bound for the radius of the extra dimensions of the order of $\SI{e-6}{m}$, which is comparable to the limits obtained from direct search of the Kaluza-Klein excitations of the graviton\cite{PDG}. Nevertheless, for higher numbers of extra dimensions we found that the constrains are a lot  more stringent. This is due to the fact that the fundamental gravitational scale, $M_*$, is related with the effective one, say $M'_{\text{pl}}$, through higher powers, reducing the dependence of the model on the size of the extra dimensions. This result provides an example of the possibility of testing non-trivial extensions of General Relativity using collider data.

A refining of the analysis can be achieved by introducing different profile functions to left and right chiralities in Eq.~\eqref{fermdecomp}, and performing the dimensional reduction analysis.\cite{PopeKK,CastilloFelisola:2012ez,Castillo-Felisola:2013jva}

This model illustrates that in the context of extra dimensions, extensions of the gravitational sector might produce interesting effects in collider phenomenology. %%Indeed, in a previous work it has been shown  that bulk gravitation could originate fermion masses%for 
For example, even the physics underlying quark masses might have a gravitational origin in the bulk.\cite{Castillo-Felisola:2013jva} Moreover, the same type of contact interaction, generates corrections to very precisely measured  electroweak observables, such as the $Z^0$ decay, and their comparison to experimental values provide additional constraints on the scales of the effective coupling constant.\cite{OCF-future2} Additionally, the structure of the contact interactions and their universality allows a new contact interaction among neutrinos, which may be important in cosmological contexts.

\section*{Acknowledgements}

We would like to thank to C. Dib, N. Neills, J.C. Helo and A. Carcamo for fruitful discussions and encourage through the realisation of this work. 

This work was supported in part by Fondecyt Projects No. 11000287 and No. 1120346, Conicyt Grant No. 21130179 and  Basal Project FS0821.

%% \appendix

%% \bibliographystyle{ws-mpla}
%% \bibliography{./References.bib}

\end{document}